\documentclass[prb,showpacs,preprintnumbers,amsmath,amssymb]{revtex4}

\usepackage{graphicx}
\usepackage{dcolumn}
\usepackage{bm}

\newcommand{\kk}[0]{{\bf k}}
\newcommand{\ko}[0]{{{\bf k}_0}}
\newcommand{\kol}[0]{{{\bf k}_0^<}}
\newcommand{\kog}[0]{{{\bf k}_0^>}}

\newcommand{\kkol}[0]{{{\bf k}+{\bf k}_0^<}}
\newcommand{\kkog}[0]{{{\bf k}+{\bf k}_0^>}}
\newcommand{\0}[0]{{\bf 0}}
\newcommand{\Bb}[0]{{\mathcal{B}}}
\newcommand{\Ee}[0]{{\mathcal{E}}}

\begin{document}

\title{Spin-flop transition accompanied with changing the type of magnetic ordering}

\author{A.\ V.\ Syromyatnikov$^{1,2}$}
\email{asyromyatnikov@yandex.ru}
\affiliation{$^1$St.\ Petersburg State University, 7/9 Universitetskaya nab., St.\ Petersburg, 199034
Russia}
\affiliation{$^2$National Research Center "Kurchatov Institute" B.P.\ Konstantinov Petersburg Nuclear Physics Institute, Gatchina 188300, Russia}

\date{\today}

\begin{abstract}

We discuss theoretically rather rear example of spin-flop transition which is accompanied with changing the type of magnetic ordering and which seemingly has not been addressed yet. We demonstrate that changing the type of magnetic ordering can manifest itself in antiferromagnetic (AF) resonance experiments as an apparent peculiar switching of the anisotropy at the transition from the easy-axis type to the easy-plane one. We argue that this kind of spin-flop transition is observed recently by K.\ Yu.\ Povarov et al., Phys.\ Rev.\ B {\bf 87}, 214402 (2013) in $\rm Cu(pz)_2(ClO_4)_2$ (where pz denotes pyrazine), one of the best realizations of spin-$\frac12$ Heisenberg AFs on square lattice having a very small anisotropy. We show that the magnetic ordering changes at the spin-flop transition in this material in the direction perpendicular to AF square planes. We examine the microscopic mechanism of such behavior in $\rm Cu(pz)_2(ClO_4)_2$ and find that dipolar forces and extremely small exchange coupling between spins from neighboring planes are responsible for it. 

\end{abstract}


\maketitle

\section{Introduction}

Spin-$\frac12$ Heisenberg antiferromagnet (AF) on square lattice has been one of the most extensively discussed models of quantum magnetism in recent three decades. Apart from its relevance to parent compounds of high temperature cuprate superconductors, \cite{monous} it provides a convenient playground for investigation of novel types of many-body phenomena, among which are quantum spin-liquid and nematic phases, novel universality classes of phase transitions, and order-by-disorder phenomena. In the most simple variant of this model with exchange coupling between only nearest-neighbor spins, the N\'eel order arises at $T=0$ which is destroyed by thermal fluctuations at any finite $T$ according to Mermin-Wagner theorem. \cite{merwag} However all practical three-dimensional (3D) realizations of this model contain weak low-symmetry interactions of relativistic nature and exchange interaction between spins from different planes which lead to a finite transition temperature $T_N$ to the N\'eel phase. In particular, the role is well known of unavoidable long-range dipolar interaction in stabilization of magnetically ordered phases in low-dimensional magnetic systems. \cite{mal,revdip}

\begin{figure}
\noindent
\includegraphics[scale=0.6]{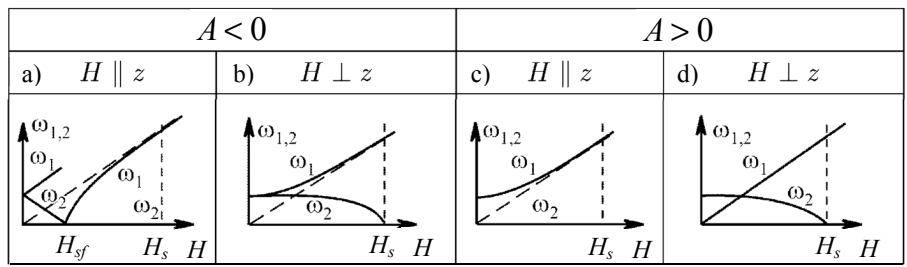}
\hfil
\caption{Sketch of normal behavior of AFR frequencies $\omega_{1,2}$ in easy-axis ($A<0$) and easy-plane ($A>0$) two-sublattice antiferromagnets in magnetic field $H$, where $A$ is the anisotropy value and $z$ is easy/hard axis. \cite{afr} $H_{sf}$ and $H_s$ are spin-flop and saturation (spin-flip) fields, respectively. 
\label{afrfig}}
\end{figure}

\begin{figure}
\noindent
\includegraphics[scale=0.6]{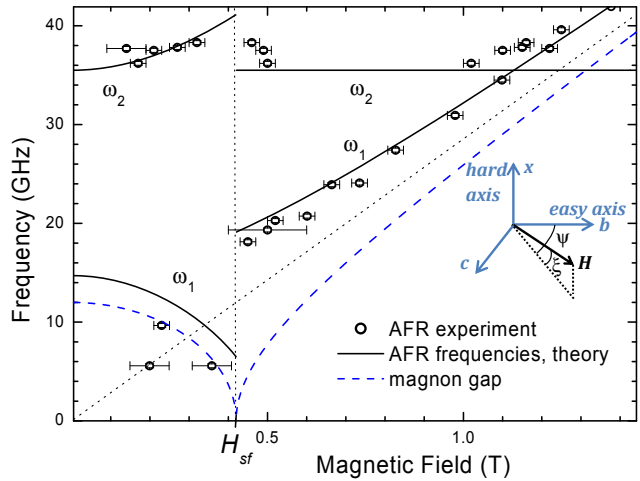}
\hfil
\caption{(Color online.) AFR data obtained in Ref.~\cite{smirnov} in $\rm Cu(pz)_2(ClO_4)_2$ at $T=1.3$~K for magnetic field directed along the easy axis inside the square planes ($\psi=\xi=0$ in the inset). It is seen from Fig.~\ref{afrfig} that the system behaves according to Fig.~\ref{afrfig}(a) and \ref{afrfig}(c) at $H<H_{sf}$ and $H>H_{sf}$, respectively. Solid lines are theoretical results obtained in the present paper using model \eqref{ham} in the first order in $1/S$. Magnon gap is also shown which is derived theoretically in the present work. 
\label{expfig}}
\end{figure}

$\rm Cu(pz)_2(ClO_4)_2$ (where pz denotes pyrazine) has been found recently to be one of the most perfect realizations of spin-$\frac12$ Heisenberg AFs on square lattice with nearest-neighbor exchange coupling constant $J\approx18.1$~K, $T_N\approx4.21$~K, and a very small easy-plane anisotropy forcing magnetic moments to lie within the square planes. \cite{pz1,pz2,pz3,fam,pzstruct,fam1,fam2} Exchange coupling constant between spins from neighboring planes does not exceed 9~mK and dipolar forces are expected to play a significant role in the interlayer coupling. \cite{fam} Recent antiferromagnetic resonance (AFR) experiment \cite{smirnov} reveals also an in-plane easy-axis anisotropy (that is an order of magnitude smaller than the easy-plane one) and a related spin-flop transition at magnetic field $H=H_{sf}\approx0.42$~T having anomalous properties. According to the common wisdom, \cite{afr} one expects four possible scenarios summarized in Fig.~\ref{afrfig} for AFR frequencies dependence on $H$ in a two-sublattice AF in the field directed along easy/hard axis. However non of them is realized in $\rm Cu(pz)_2(ClO_4)_2$ as it is seen in Fig.~\ref{expfig}: $\rm Cu(pz)_2(ClO_4)_2$ behaves as an easy-axis AF (see Fig.~\ref{afrfig}(a)) and as easy-plane AF (see Fig.~\ref{afrfig}(c)) at $H<H_{sf}$ and $H>H_{sf}$, respectively. Then, the experimentally obtained picture looks as if the in-plane anisotropy changes at the spin-flop transition from the easy-axis type to the easy-plane one. The origin of this peculiar behavior has not been clarified yet. As it is argued in Ref.~\cite{smirnov}, a magnetoelastic mechanism cannot be responsible for it.

\begin{figure}
\noindent
\includegraphics[scale=0.5]{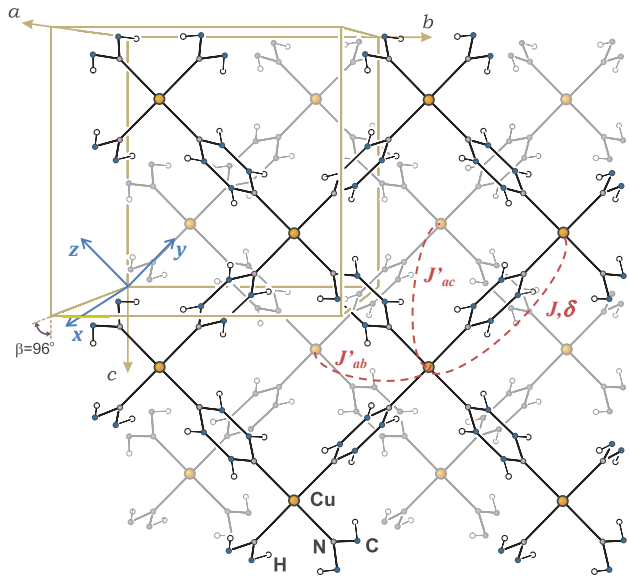}
\hfil
\caption{(Color online.) $\rm Cu(pz)_2(ClO_4)_2$ structure obtained in crystallographic measurements \cite{pzstruct} (this figure is adopted from Ref.~\cite{smirnov}). Two layers are shown each containing a square magnetic lattice. ClO$_4$ complexes are not shown for clarity. All symbols are faded out corresponding to the lower layer. Cartesian coordinate system $xyz$ is also shown which is used in our consideration. Parameters of model \eqref{ham} (exchange coupling constants $J$, $J'_{ab}$, $J'_{ac}$, and exchange anisotropy $\delta$) are also shown.
\label{pzfig}}
\end{figure}

In the present paper, we propose a microscopic model and describe quantitatively low-temperature experimental data reported before for $\rm Cu(pz)_2(ClO_4)_2$. The model Hamiltonian we discuss contains Heisenberg spin coupling, anisotropic easy-plane exchange interaction, and unavoidable dipolar forces. In particular, we demonstrate in Sec.~\ref{sfdip}, where the classical ground state energy is analyzed, that the in-plane easy-axis anisotropy obtained in Ref.~\cite{smirnov} originates from dipolar interaction and a small departure of the crystal structure from the tetragonal one (in particular, from small deviation of the angle $\beta$ from $90^\circ$ depicted in Fig.~\ref{pzfig}). We show that dipolar forces lead to changing at the spin-flop transition of the magnetic ordering in the direction perpendicular to square planes if inter-plane exchange coupling is sufficiently small. This changing of the magnetic ordering takes place at $\xi<\xi_c$ and $\psi<\psi_c$, where $\xi$ and $\psi$ are angles determining the field orientation (see inset in Fig.~\ref{expfig}). We find expressions for critical angles $\xi_c$ and $\psi_c$.

Magnon spectrum, AFR frequencies and ground state energy are derived in Sec.~\ref{magnon} in the first order in $1/S$. We demonstrate that due to the changing of the magnetic ordering at the spin-flop transition, the lower AFR frequency does not coincide with the gap $\Delta$ in lower magnon branch (as usual, $\Delta$ vanishes at the spin-flop transition) and AFR frequencies behave in this model as those experimentally obtained in $\rm Cu(pz)_2(ClO_4)_2$ (see Fig.~\ref{expfig}). In Sec.~\ref{compexp}, we use analytical expressions obtained to fit available low-temperature experimental data and to extract parameters of the microscopic model. In particular, we conclude that the inter-plane exchange coupling cannot exceed a value of the order of $10^{-3}$~K in $\rm Cu(pz)_2(ClO_4)_2$. Then, it is the tiny inter-plane exchange coupling that makes possible the peculiar behavior of $\rm Cu(pz)_2(ClO_4)_2$ in magnetic field which is governed by dipolar forces.

We summarize our results in Sec.~\ref{conc}. To provide an intuitively clear example of a system showing the anomalous spin-flop transition similar to that obtained in $\rm Cu(pz)_2(ClO_4)_2$, we discuss also in Sec.~\ref{conc} a phenomenological model of a layered two-sublattice AF having a hierarchy of small anisotropic spin interactions. These interactions lead to changing the type of magnetic ordering at the spin-flop transition in the direction perpendicular to AF planes. Then, the effect of long-range dipolar interaction in the microscopic model is simulated by the hierarchy of anisotropic short-range spin interactions in the phenomenological model.

Our consideration can be relevant also to members of a family of recently synthesized \cite{fam} two-dimensional (2D) spin-$\frac12$ Heisenberg AFs to which $\rm Cu(pz)_2(ClO_4)_2$ is a prototype.

\section{Microscopic model for $\rm Cu(pz)_2(ClO_4)_2$. Classical ground state analysis.}
\label{sfdip}

In this section, we discuss classical ground-state properties of a model which we use in Sec.~\ref{compexp} to describe quantitatively low-temperature experimental data reported before for $\rm Cu(pz)_2(ClO_4)_2$. The model Hamiltonian has the form
\begin{equation}
\label{ham}
\mathcal{H} = \frac{1}{2}\sum_{l\neq m}
\left(
J_{lm}\delta_{\alpha \gamma} - Q_{lm}^{\alpha \gamma}
\right) S_l^\alpha S_{m}^\gamma
-
\delta\sum_{\langle l,m \rangle} S_l^x S_{m}^x
-
{\bf H}\sum_l {\bf S}_l,
\end{equation}	
where the summation over repeated Greek letters is implied, $\langle l,m \rangle$ denote nearest-neighbor couples of spins in square planes, $\delta>0$ is the value of easy-plane anisotropy, $x$ is the hard axis (see Fig.~\ref{pzfig}), the first term describes the short-range exchange and long-range dipolar interaction between spins, dipolar tensor $Q$ has the form
\begin{eqnarray}
\label{dip_forces}
Q_{lm}^{\alpha \gamma} &=& \frac{\omega_0}{4\pi}\frac{3R_{lm}^\alpha R_{lm}^\gamma-\delta_{\alpha \gamma}R_{lm}^2}{R_{lm}^5},\\
\label{w0}
\omega_0 &=& 4\pi \frac{(g\mu_B)^2}{v_0},
\end{eqnarray}
$v_0$ is the unit cell volume, and $\omega_0$ is the characteristic dipolar energy. $g$-factor is anisotropic in $\rm Cu(pz)_2(ClO_4)_2$: $g=g_{bc}=2.04(3)$ for in-plane components of magnetic moments and $g=g_x=2.25(5)$ for components perpendicular to square planes. \cite{smirnov,fam} Then, $\omega_0$ can possess three slightly different values in Eq.~\eqref{dip_forces} depending on $\alpha$ and $\gamma$ that is taken into account in calculations below. Notice also that Dzyaloshinsky-Moriya interaction between spins is forbidden in $\rm Cu(pz)_2(ClO_4)_2$ by the crystal symmetry. \cite{pz3}

To calculate accurately dipolar tensor $Q$, one has to take into account the crystal structure of $\rm Cu(pz)_2(ClO_4)_2$ obtained in Refs.~\cite{pzstruct,fam} and shown in Fig.~\ref{pzfig}. The unit cell containing four Cu$^{2+}$ magnetic atoms (two atoms from two neighboring $bc$ planes) has parameters $a=13.9276(3)$~\AA, $b=9.7438(2)$~\AA, and $c=9.7871(2)$~\AA. \cite{fam} There are right angles between $b$ and $c$ axes as well as between $a$ and $b$ ones. The angle $\beta=96.924(1)^\circ$ between $a$ and $c$ axes differs slightly from $90^\circ$. \cite{fam} We neglect below for simplicity the tiny difference between lattice parameters $b$ and $c$. Thus, we assume that Cu$^{2+}$ ions are arranged in perfect square lattice in $bc$ planes. On the other hand, we show below that the small departure of angle $\beta$ from $90^\circ$ is responsible for the in-plane anisotropy observed in Ref.~\cite{smirnov}. Adjacent layers are shifted relative to each other by a half of a period. The coordinate system describing the magnetic subsystem can be built on basis vectors 
\begin{eqnarray}
\label{evec}
	{\bf e}_1 &=& \frac{b}{\sqrt2}(0,0,1),\nonumber\\
	{\bf e}_2 &=& \frac{b}{\sqrt2}(0,1,0),\\
	{\bf e}_3 &=& \frac{1}{2\sqrt2}
(\sqrt2a \sin\beta, b - a\cos\beta,b - a\cos\beta) \nonumber
\end{eqnarray}
so that ${\bf R}_{lm}=n_1{\bf e}_1+n_2{\bf e}_2+n_3{\bf e}_3$ in Eq.~\eqref{dip_forces}, where $n_{1,2,3}$ are integer and components of ${\bf e}_{1,2,3}$ are given in the Cartesian coordinate system whose $y$ and $z$ axes are parallel to square edges inside $bc$ planes and $x$ axis is perpendicular to $bc$ planes (see Fig.~\ref{pzfig}). Notice that the unit cell built on vectors ${\bf e}_{1,2,3}$ contains one magnetic atom. In particular, the characteristic dipolar energy given by Eq.~\eqref{w0} is equal approximately to 0.1~K at $g=g_{bc}$ in $\rm Cu(pz)_2(ClO_4)_2$. A vector $\kk$ in the reciprocal space is assumed below to have the form $\kk = (k_1,k_2,k_3) = k_1{\bf b}_1+k_2{\bf b}_2+k_3{\bf b}_3$, where ${\bf b}_1=[{\bf e}_2\times{\bf e}_3]/v_0$, ${\bf b}_2=[{\bf e}_3\times{\bf e}_1]/v_0$, and ${\bf b}_3=[{\bf e}_1\times{\bf e}_2]/v_0$.

\subsection{Classical ground state at $H=0$. In-plane easy-axis anisotropy.}

Let us try the classical ground state magnetic ordering at $H=0$ in the form
\begin{equation}
\label{order}
{\bf S}_j  = S 
\left({\bf e}_1 \cos (\alpha + \ko{\bf R}_j) + {\bf e}_2 \sin(\alpha + \ko{\bf R}_j)\right) 
\end{equation}
where $\ko$ is a vector of magnetic structure and we assume that all magnetic moments lie within $bc$ planes due to the easy-plane anisotropy. Angle $\alpha$ and $\ko$ should be found by minimization of the classical energy $E_{cl}$ which is obtained by substitution of Eq.~\eqref{order} to Eq.~\eqref{ham}:
\begin{equation}
\label{ecl}
	\frac{2}{NS^2} E_{cl} = J_\ko + J'_\ko - \left(Q_\ko^{bb}\cos^2\alpha+Q_\ko^{cc}\sin^2\alpha\right),
\end{equation}
where $N$ is the number of spins in the system, $Q_\kk^{\alpha \beta}=\sum_l Q_{l m}^{\alpha \beta}\exp(i\kk {\bf R}_{l m})$, dipolar tensor components are taken in the Cartesian coordinate system shown in Fig.~\ref{pzfig} with axes $x$, $b$, and $c$, $J'_\kk=\sum_l J'_{l m}\exp(i\kk {\bf R}_{l m})$ and $J_\kk=\sum_l J_{l m}\exp(i\kk {\bf R}_{l m})$ are Fourier transformation of the inter- and intra-plane exchange couplings, respectively. Strong in-plane nearest-neighbor AF exchange coupling $J>0$ fixes two components of $\ko=k_{01}{\bf b}_1+k_{02}{\bf b}_2+k_{03}{\bf b}_3$:
\begin{equation}
\label{k12}
k_{01}=k_{02}=\pi.
\end{equation}
The first term on the right-hand side of Eq.~\eqref{ecl} reaches its minimum of $-4J$ in this case. Notice that the third term in Eq.~\eqref{ecl} does not effect $k_{01}$ and $k_{02}$ because $Q_\kk^{cc}$ and $Q_\kk^{bb}$ are quadratic functions of momenta near points $\kk=\pi{\bf b}_1+\pi{\bf b}_2+k_3{\bf b}_3$ at any given $k_3$. It is seen from Eq.~\eqref{order} that Eq.~\eqref{k12} implies the N\'eel ordering in square planes.

\begin{figure}
\noindent
\includegraphics[scale=0.5]{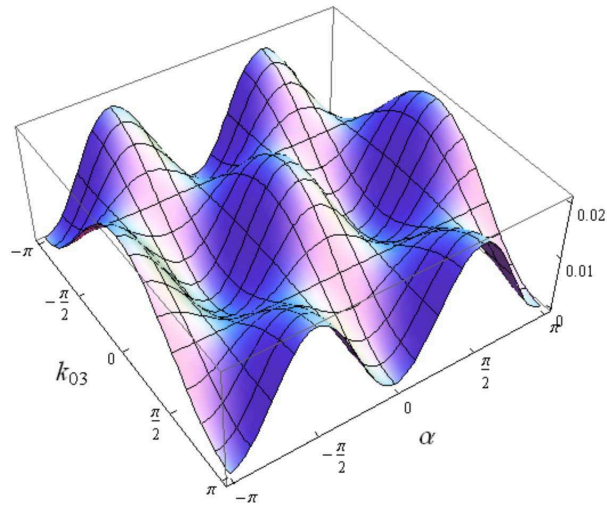}
\hfil
\caption{(Color online.) Dependence on $\alpha$ and $k_{03}$ of the third term on the right-hand side of Eq.~\eqref{ecl} for the classical ground-state energy at $H=0$. Among two inequivalent local minima at $\alpha=0$, $k_{03}=\pi$ and $\alpha=\pi/2$, $k_{03}=0$ corresponding to the direction of the staggered magnetization along $b$ and $c$ axes, respectively, the former one has lower energy due to the departure of $\beta$ from $90^\circ$ (see Fig.~\ref{pzfig}). This is the origin of the easy-axis in-plane anisotropy obtained experimentally in Ref.~\cite{smirnov}.
\label{eclfig}}
\end{figure}

We have calculated dipolar tensor components numerically using dipolar sums computation technique \cite{cohen} and found that the dependence of $Q_\ko^{cc}$ and $Q_\ko^{bb}$ on $k_{03}$ can be approximated very accurately as follows when Eq.~\eqref{k12} holds: 
\begin{equation}
\label{qq}
\begin{aligned}
Q_\ko^{cc} &\approx \frac12 \left(Q_\kog^{cc}+Q_\kol^{cc} + \left(Q_\kog^{cc}-Q_\kol^{cc}\right)\cos k_{03}\right),\\
Q_\ko^{bb} &\approx \frac12 \left(Q_\kog^{bb}+Q_\kol^{bb} + \left(Q_\kog^{bb}-Q_\kol^{bb}\right)\cos k_{03}\right),
\end{aligned}
\end{equation}
where $\kol = (\pi,\pi,\pi)$, $\kog = (\pi,\pi,0)$, $Q_\kog^{cc}\approx-0.01733\omega_0$, $Q_\kol^{cc}\approx-0.19472\omega_0$, $Q_\kog^{bb}\approx-0.2095\omega_0$, and $Q_\kol^{bb}\approx-0.00384\omega_0$. Dependence on $\alpha$ and $k_{03}$ is shown in Fig.~\ref{eclfig} of the third term on the right-hand side of Eq.~\eqref{ecl} which is found using Eqs.~\eqref{qq}. It is seen that there are two non-equivalent local minima at 
\begin{equation}
\label{min}
k_{03}=\pi, \quad \alpha=0
\end{equation}
and at $k_{03}=0$, $\alpha=\pi/2$ corresponding to the direction of the staggered magnetization along $b$ and $c$ axes, respectively. If $\beta$ was equal to $90^\circ$, energies would be the same in these minima. However the small departure of $\beta$ from $90^\circ$ makes the point \eqref{min} to be the absolute minimum of the energy. 

AF coupling of nearest-neighbor spins from adjacent layers does not change qualitatively this picture. As soon as adjacent layers are shifted relative to each other by a half of a period in $\rm Cu(pz)_2(ClO_4)_2$, \cite{pzstruct} a magnetic ion within a layer is equidistant from four ions in the adjacent layer. If we assume that a spin from a layer is coupled by exchange interaction to all four its nearest neighbors from an adjacent layer, the exchange coupling constant being $J'$, the exchange interlayer coupling would cancel in the classical limit: $J'_\ko=0$ in Eq.~\eqref{ecl} at any $k_{03}$. However the interlayer interaction pathway along $ac$ diagonal is shorter in $\rm Cu(pz)_2(ClO_4)_2$ than that along $ab$ diagonal. \cite{pz3} Then, we introduce two exchange AF coupling constants, $J'_{ab}$ and $J'_{ac}>J'_{ab}$, for interaction between a given spin and its two neighbors from an adjacent layer along $ab$ and $ac$ diagonals, respectively (see Fig.~\ref{pzfig}). The assumption about the positiveness of $J'_{ac}$ is confirmed below in comparison of the theory with experiment. In this case $J'_\ko=4(J'_{ac}-J'_{ab})>0$ and $J'_\ko=-4(J'_{ac}-J'_{ab})<0$ if $k_{03}=0$ and $k_{03}=\pi$, correspondingly. Then, the interlayer interaction further lowers the energy of minimum \eqref{min} and increases the energy of another local minimum (see Eq.~\eqref{ecl}).

To summarize this subsection, we obtain that i) in accordance with the experimental finding of Ref.~\cite{smirnov}, $b$ axis is the easy direction inside the easy $bc$ plane ($\alpha=0$) and ii) in agreement with neutron diffraction experiment \cite{pz3}, magnetic moments arrange antiferromagnetically in adjacent layers along $ac$ diagonal at $H=0$ ($k_{03}=\pi$).

\subsection{Classical ground state at field parallel to easy axis}

When the field is directed along the easy $b$ axis, the energy of the collinear AF phase 
\begin{equation}
\label{eclcol}
\frac1N E_{cl}^{col} = \frac{S^2}{2} \left(-J_\0 + J'_\ko - Q_\ko^{bb}\right), 
\end{equation}
would be lower than the energy $E_{cl}^{sf}$ of the spin-flopped phase at $H<H_{sf}$, where $J_\0=4J$. To find $E_{cl}^{sf}$ and $H_{sf}$, we assume that the canted AF ordering has the form at $H>H_{sf}$
\begin{equation}
\label{ordersf}
{\bf S}_j = S \left( \hat b \cos\theta + \hat c e^{i\ko{\bf R}_j} \sin\theta \right),
\end{equation}
where $\hat b$ and $\hat c$ are unit vectors directed along corresponding axes and $e^{i\ko{\bf R}_j}$ is equal to $+1$ and $-1$ on sites belonging to different AF sublattices. Substituting Eq.~\eqref{ordersf} to Eq.~\eqref{ham} and minimizing the energy with respect to $\theta$, we obtain in the leading orders in small parameters $\omega_0$, $J'$, and $H$
\begin{eqnarray}
\label{eclsf}
\frac1N E_{cl}^{sf} &\approx& \frac{S^2}{2} \left(-J_\0 + J'_\ko - Q_\ko^{cc}\right) - \frac{H^2}{4J_\0},\\
\label{theta}
\cos\theta &\approx& \frac{H}{2J_\0S}.
\end{eqnarray}
It is seen from Eqs.~\eqref{qq} that the third term in the brackets in Eq.~\eqref{eclsf} reaches its minimum at $k_{03}=0$ whereas the second term 
$J'_\ko = 4(J_{ac}'-J_{ab}')\cos k_{03}$ 
has a minimum at $k_{03}=\pi$. The third term wins in this competition provided that the interlayer coupling is sufficiently small:
\begin{equation}
\label{ineq}
J_{ac}'-J_{ab}' < \frac18\left(Q^{cc}_\kog - Q^{cc}_\kol\right)
\approx 0.022\omega_0 \approx 0.002~{\rm K}.
\end{equation}
Importantly, the vector of magnetic ordering changes at the spin-flop transition if Eq.~\eqref{ineq} holds:
\begin{equation}
\label{ko}
\ko = \left\{
\begin{aligned}
\kol = (\pi,\pi,\pi), & \qquad H<H_{sf},\\
\kog = (\pi,\pi,0), & \qquad H>H_{sf}.\\
\end{aligned}
\right.
\end{equation}
As we demonstrate below, it is the scenario of $\ko$ switching at $H=H_{sf}$ that is consistent with AFR experimental data reported in Ref.~\cite{smirnov}.

Comparing energies \eqref{eclcol} and \eqref{eclsf} and taking into account Eq.~\eqref{ko}, one finds for the spin-flop field value
\begin{equation}
\label{hsf}
H_{sf} = S \sqrt{ 2J_\0 \left( 8(J_{ac}'-J_{ab}') + Q^{bb}_\kol - Q^{cc}_\kog \right) }.
\end{equation}

\subsection{Classical ground state in inclined magnetic field. Critical angles.}

In this subsection, we discuss the classical ground-state energy both numerically and analytically assuming that the field is directed arbitrary relative to the easy axis. 
We consider below in some details two most representative cases of $\bf H$ lying in $bc$ and $xb$ planes. All analytical expressions found below are in very good agreement with results of the corresponding numerical consideration of the ground-state energy.

Let us assume that the magnetic field is directed by an angle $\psi$ with respect to the easy axis and that it lies in $bc$ plane (i.e., $\psi\ne0$ and $\xi=0$ in the inset of Fig.~\ref{expfig}). Sublattices magnetizations lie within $bc$ plane at any $H$ due to the easy-plane anisotropy. At large enough field $SJ_\0\gg H\gg H_{sf}$, when AF sublattices are nearly perpendicular to the field, the ground state energy has the form (cf.\ Eq.~\eqref{eclsf})
\begin{equation}
\label{eclsfin}
\frac1N E_{cl} \approx \frac{S^2}{2} 
\left(
-J_\0 + J'_\ko - Q_\ko^{cc}
+ \left(Q_\ko^{cc}-Q_\ko^{bb}\right) \sin^2\psi
\right) - \frac{H^2}{4J_\0}
\end{equation}
Comparing energies of configurations with $\ko=\kol$ and $\ko=\kog$, one finds using Eqs.~\eqref{qq} and \eqref{eclsfin} that the latter configuration is more energetically favorable (so that the first-order transition takes place at some $H\sim H_{sf}$ from $\ko=\kol$ to $\ko=\kog$) if
\begin{equation}
\label{ineqin}
J_{ac}'-J_{ab}' < \frac18
\left( 
Q^{cc}_\kog - Q^{cc}_\kol 
+ 
\left( Q_\kol^{cc}-Q_\kol^{bb} - Q_\kog^{cc}+Q_\kog^{bb} \right) \sin^2\psi 
\right).
\end{equation}
This inequality is a counterpart of Eq.~\eqref{ineq}. It is seen from Eqs.~\eqref{qq} that coefficient is negative before $\sin^2\psi$ in Eq.~\eqref{ineqin}. Then, inequality \eqref{ineqin} breaks (and the transition does not occur) at $\psi>\psi_c$, where
\begin{equation}
\label{psic}
\sin\psi_c = \sqrt{\frac{ Q^{cc}_\kog - Q^{cc}_\kol - 8\left(J_{ac}'-J_{ab}' \right) }{ Q_\kog^{cc}-Q_\kog^{bb} - Q_\kol^{cc}+Q_\kol^{bb}}}.
\end{equation}

\begin{figure}
\noindent
\includegraphics[scale=0.7]{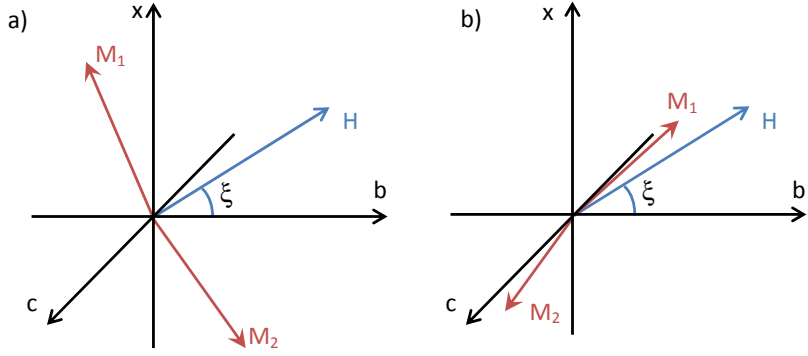}
\hfil
\caption{(Color online.) Possible configurations of sublattices magnetizations ${\bf M}_{1,2}$ when the field $\bf H$ lies in $xb$ plane and $SJ_\0\gg H\gg S\sqrt{J_\0\delta_\0}$. ${\bf M}_{1,2}$ lie in $xb$ plane and they are slightly out of $bc$ plane in configurations shown on panel a) and b), respectively. Energy of configuration a) is lower than that of b) at $\xi>\xi_c$ which are given by Eqs.~\eqref{xic} and \eqref{xicr}. As soon as the vector of magnetic structure $\ko$ is equal to $\kol$ and $\kog$ in configurations a) and b), respectively, the switching takes place of $\ko$ from $\kol$ to $\kog$ at some field value if $\xi<\xi_c$.
\label{axes}}
\end{figure}

Let us direct now $\bf H$ by angle $\xi$ relative to the easy axis and assume that the field lies within $xb$ plane (i.e., $\psi=0$ and $\xi\ne0$ in the inset of Fig.~\ref{expfig}). Particular analytical consideration supported by corresponding numerical calculations shows that two spin configurations compete at $SJ_\0\gg H\gg S\sqrt{J_\0\delta_\0}$ which are presented in Figs.~\ref{axes}(a) and \ref{axes}(b) and which energies have the form
\begin{eqnarray}
\label{ea}
\frac1N E_{cl}^{(a)} &\approx& \frac{S^2}{2} 
\left(
-J_\0 + J'_\ko - Q_\ko^{xx}
+ \left(Q_\ko^{xx}-Q_\ko^{bb}\right) \sin^2\xi
+ \delta_\0 \cos^2\xi
\right) - \frac{H^2}{4J_\0},\\
\label{eb}
\frac1N E_{cl}^{(b)} &\approx& \frac{S^2}{2} \left(-J_\0 + J'_\ko - Q_\ko^{cc}\right) - \frac{H^2}{4J_\0},
\end{eqnarray}
where $\delta_\0=4\delta$ and $E_{cl}^{(b)}\approx E_{cl}^{sf}$ given by Eq.~\eqref{eclsf}. Using Eqs.~\eqref{qq} and equality $Q_\ko^{xx} = -(g_x/g_{bc})^2(Q_\ko^{bb}+Q_\ko^{cc})$, one concludes that configurations with $\ko=\kol$ and $\ko=\kog$ minimize energies $E_{cl}^{(a)}$ and $E_{cl}^{(b)}$, respectively. Then, the transition takes place from $\ko=\kol$ to $\ko=\kog$ if $E_{cl}^{(b)}<E_{cl}^{(a)}$ that implies $\xi<\xi_c$, where
\begin{equation}
\label{xic}
\cos\xi_c = \sqrt{
\frac{8 (J_{ac}-J_{ab})-Q_\kog^{cc} + Q_\kol^{bb}}{\delta_\0 - Q_\kol^{xx} + Q_\kol^{bb} }
}.
\end{equation}

Notice that critical angles $\psi_c$ and $\xi_c$ of the considered spin-flop transition can reach several tenths degrees (in particular, $\psi_c\approx10^\circ$ and $\xi_c\approx30^\circ$ in $\rm Cu(pz)_2(ClO_4)_2$ \cite{smirnov}). In contrast, critical angles in common spin-flop transitions are usually of several degrees because they are governed by small ratio of anisotropy value to exchange coupling constant and/or small ratios of anisotropies of different orders (cf.\ Eqs.~\eqref{psic} and \eqref{xic}). \cite{bogdan}

\section{Magnon spectrum in $\rm Cu(pz)_2(ClO_4)_2$ and quantum corrections to observables}
\label{magnon}

We analyze in this section in detail magnon spectrum when the field is directed along the easy axis. Corresponding consideration for arbitrary field direction can be carried out accordingly. However, the results are cumbersome and we do not present them below.

\subsection{Magnon spectrum in collinear phase. $H<H_{sf}$.}

Assuming that the magnetic field is directed along the easy $b$ axis, it is convenient to represent spin components as follows: 
${\bf S}_j=S_j^x\hat{x}+(S_j^c\hat{c}+S_j^b\hat{b})\exp(i{\bf R}_j\kol)$, where $\hat{x}$, $\hat{b}$, and $\hat{c}$ are unit vectors directed along corresponding axes (see Fig.~\ref{pzfig}). We use below the Holstein-Primakoff spin representation having the form
\begin{eqnarray}
\label{hp}
S_j^x &\approx& \sqrt{\frac{S}{2}}\left(  a_j+a_j^\dagger-\frac{a^\dagger_j a_j^2 + (a^\dagger_j)^2 a_j}{4S}
\right), \nonumber\\
S_j^c &\approx& -i\sqrt{\frac{S}{2}}\left(   a_j-a_j^\dagger-\frac{a^\dagger_j a_j^2 - (a^\dagger_j)^2 a_j}{4S}
\right), \\
S_j^b &=& S-a^\dagger_j a_j. \nonumber
\end{eqnarray}
Substituting Eqs.~\eqref{hp} to Eq.~\eqref{ham}, one obtains for the Hamiltonian $\mathcal{H}=E_{cl}+\sum_{i=1}^6\mathcal{H}_i$, where $E_{cl}$ is given by Eq.~\eqref{eclcol} and $\mathcal{H}_i$ denote terms containing products of $i$ operators $a^\dagger$ and $a$. In particular, one has
\begin{eqnarray}
\label{h1}
\frac{1}{\sqrt N}\mathcal{H}_1 &=& -S\sqrt{\frac{S}{2}}Q^{xb}_\kol(a^{}_\kol+a^\dagger_\kol)+i S\sqrt{\frac{S}{2}}Q^{bc}_\kol(a_\0-a^\dagger_\0),\\
\label{h2}
\mathcal{H}_2 &=& \sum_{\bf k}
\left( 
E_{\bf k}a^\dagger_{\bf k}a^{}_{\bf k} + \frac{B_{\bf k}}{2}\left(a^{}_{\bf k}a^{}_{-{\bf k}}+a^\dagger_{\bf k}a^\dagger_{-{\bf k}}\right)
+(\mathcal{E}_{\bf k} + H)a^\dagger_{{\bf k}+{\bf k}_0}a^{}_{\bf k}
+ \frac{\mathcal{B}_{\bf k}}{2} a^{}_{{\bf k}}a^{}_{-{\bf k}+{\bf k}_0}
+ \frac{\mathcal{B}_{\bf k}^*}{2} a^\dagger_{{\bf k}}a^\dagger_{-{\bf k}+{\bf k}_0}
\right),
\end{eqnarray}
where $\mathcal{H}_1=0$ because $Q^{xb}_\kol=Q^{bc}_\kol=0$,
\begin{subequations}
\label{h2coef}
\begin{eqnarray}
E_\kk & = &  \frac{S}2\left( \tilde J_\kk + \tilde J_\kkol - 2 \tilde J_\kol \right) 
- \frac{S}{2}\left(Q^{xx}_\kk+Q^{cc}_\kkol-2Q^{bb}_\kol  + \delta_\kk \right),\\
B_\kk & = &  \frac{S}2\left( \tilde J_\kk - \tilde J_\kkol \right) 
- \frac{S}{2}\left(Q^{xx}_\kk-Q^{cc}_\kkol + \delta_\kk \right),\\
\mathcal{E}_\kk & = & i\frac{S}{2}\left( Q_\kkol^{xc}-Q^{xc}_\kk\right),\\
\mathcal{B}_\kk & = & i\frac S2 \left( Q_\kkol^{xc}+Q^{xc}_\kk\right),
\end{eqnarray}
\end{subequations}
and $\tilde J_\kk = J_\kk + J'_\kk$. The bilinear part of the Hamiltonian \eqref{h2} can be analyzed as it is done, e.g., in Refs.~\cite{batalov,batalov2}. One obtains that the magnon spectrum has two branches whose energies $\epsilon_{\kk}^\pm$ have the form (cf.\ Eqs.~(14) and (15) in Ref.~\cite{batalov})
\begin{eqnarray}
\label{specsw}
\left(\epsilon_{\kk}^\pm\right)^2 
&=& 
\frac12 \left(E_\kk^2+E_\kkol^2-B_\kk^2-B_\kkol^2+2\Bb_\kk^2-2\Ee_\kk^2 + 2H^2 \right)
\pm 
\sqrt{d_\kk},\nonumber\\
d_\kk &=& \frac14 \left(E_\kk^2+E_\kkol^2-B_\kk^2-B_\kkol^2+2\Bb_\kk^2-2\Ee_\kk^2 + 2H^2 \right)^2
+ 4H^2 \left(E_\kk E_\kkol + \Ee_\kk^2\right)\\
&&{}
- \left( (E_\kk+B_\kk)(E_\kkol-B_\kkol) + (\Ee_\kk-\Bb_\kk)^2 + H^2 \right) 
\left( (E_\kk-B_\kk)(E_\kkol+B_\kkol) +(\Ee_\kk+\Bb_\kk)^2 + H^2 \right).\nonumber
\end{eqnarray}
It is easy to show using Eqs.~\eqref{h2coef} and \eqref{specsw} that $\epsilon_{\kk}^\pm$ are invariant under replacement of $\kk$ by $\kk+\kol$. We have at $H\sim H_{sf}$ from Eqs.~\eqref{specsw} for energies of uniform modes which are measured in AFR experiments
\begin{eqnarray}
\label{o1}
\omega_1^{col}(H) &=& \epsilon_{\kk=\0}^- = \sqrt{ \omega_1(0)^2 - H^2 },\\
\label{o10}
\omega_1^{col}(0) &=& S\sqrt{ 2J_\0 \left( Q^{bb}_\kol - Q^{cc}_\kol \right) },\\
\label{o2}
\omega_2^{col}(H) &=& \epsilon_{\kk=\0}^+ = \sqrt{ \omega_2(0)^2 + 3H^2 },\\
\label{o20}
\omega_2^{col}(0) &=& S\sqrt{ 2J_\0 \delta_\0 },
\end{eqnarray}
where $J_\0=4J$ and $\delta_\0=4\delta$ in our model and we assume that $J\gg \delta,\omega_0,H,J'$ and $\delta\gg\omega_0$. Importantly, due to the switching \eqref{ko} of $\ko$ at the spin-flop transition, the minimum of the lower magnon branch $\epsilon_\kk^-$ is situated not at $\kk=\0$ (or, equivalently, not at $\kk=\kol$) but at $\kk=\kog$. Then, one has from Eqs.~\eqref{specsw} for the gap $\Delta$ in the lower branch (cf.\ Eq.~\eqref{o1})
\begin{equation}
\label{gapcol}
\Delta_{col} = \epsilon_{\kk=\kog}^- = \sqrt{ H_{sf}^2 - H^2 },
\end{equation}
where $H_{sf}$ is given by Eq.~\eqref{hsf}. It is seen from Eq.~\eqref{gapcol} that (quite expectedly) the gap vanishes at $H=H_{sf}$. In contrast, $\omega_1$ remains finite at $H=H_{sf}$ as soon as inequality \eqref{ineq} holds.

Thus, we stress one more time that in contrast to the common situation lower AFR frequency does not coincide with the magnon gap in the considered system.

\subsection{Magnon spectrum in spin-flopped phase. $H>H_{sf}$.}

The magnon spectrum in the spin-flopped phase can be calculated in much the same way as it is done above for the collinear phase. We use Holstein-Primakoff transformation \eqref{hp} and represent spin components as
\begin{equation}
{\bf S}_j = S_j^x\hat{x}
+
(S_j^y\hat{c}+S_j^z\hat{b})\cos\theta
+
(S_j^y\hat{b}-S_j^z\hat{c})e^{i{\bf R}_j\kog}\sin\theta.
\end{equation}
As a result, one leads to the bilinear part of the Hamiltonian \eqref{h2}, where $H$ should be discarded. In calculations performed in the leading orders in small parameters $\delta,\omega_0,J'$ and at $H\sim H_{sf}\ll SJ$, one can use Eq.~\eqref{theta} for $\theta$ and the following expressions for coefficients in ${\cal H}_2$ (cf.\ Eqs.~\eqref{h2coef}):
\begin{subequations}
\label{h2coefsf}
\begin{eqnarray}
E_\kk & = &  S\left( \tilde J_\kk - \tilde J_\kog + \frac12\left( \tilde J_\kkog -\tilde J_\kk \right) \sin^2\theta \right) 
- \frac{S}{2}\left(Q^{xx}_\kk+Q^{bb}_\kkog-2Q^{cc}_\kog  + \delta_\kk \right),\\
B_\kk & = &  \frac{S}2\left( \tilde J_\kk - \tilde J_\kkog \right) \sin^2\theta
- \frac{S}{2}\left(Q^{xx}_\kk-Q^{bb}_\kkog + \delta_\kk \right),\\
\mathcal{E}_\kk & = & i\frac{S}{2}\left( Q_\kkog^{xb}-Q^{xb}_\kk\right),\\
\mathcal{B}_\kk & = & i\frac S2 \left( Q_\kkog^{xb}+Q^{xb}_\kk\right).
\end{eqnarray}
\end{subequations} 
One leads to Eqs.~\eqref{specsw} for the magnon spectrum, where $H$ should be discarded, $\kol$ should be replaced by $\kog$, and Eqs.~\eqref{h2coefsf} should be adopted. It is easy to show that the spectrum is invariant under replacement of $\kk$ by $\kk+\kog$ (cf.\ the collinear phase). In particular, one obtains for AFR frequencies in the spin-flopped phase in the leading order in small parameters
\begin{eqnarray}
\label{o1sf}
\omega_1^{sf}(H) &=& \epsilon_{\kk=\0}^- = \sqrt{ \tilde\omega_1^2 + H^2 },\\
\label{o1sf0}
\tilde\omega_1 &=& S\sqrt{ 2J_\0 \left( Q^{cc}_\kog - Q^{bb}_\kog \right)},\\
\label{o2sf}
\omega_2^{sf}(H) &=& \epsilon_{\kk=\0}^+ = S\sqrt{ 2J_\0 \delta_\0  }=\omega_2^{col}(0).
\end{eqnarray}

Similar to the collinear phase, the minimum of $\epsilon_\kk^-$ situates not at $\kk=\0$ (or, equivalently, not at $\kk=\kog$) but at $\kk=\kol$ so that the gap value has the form
\begin{equation}
\label{gapsf}
\Delta_{sf} = \epsilon_{\kk=\kol}^- = \sqrt{H^2 - H_{sf}^2 },
\end{equation}
where $H_{sf}$ is given by Eq.~\eqref{hsf}. As in the collinear phase, the gap vanishes at $H=H_{sf}$ in contrast to $\omega_1$ that remains finite at $H=H_{sf}$ (see Eq.~\eqref{o1sf}).

\subsection{Quantum and thermal corrections to observables}
\label{qcor}

Quantum fluctuations have to be taken into account in the considered spin-$\frac12$ quasi-2D system. We show below that first $1/S$ corrections give noticeable contributions to the results obtained above not changing, however, qualitatively the observed physical picture.

Let us start with first $1/S$-corrections to ground state energies \eqref{eclcol} and \eqref{eclsf} which are determined by \cite{batalov2}
\begin{equation}
\langle {\cal H}_2 \rangle = 
\frac14\sum_{\bf k} 
\left(
\epsilon_\kk^+ \left( 1 + 2N(\epsilon_\kk^+) \right) 
+ 
\epsilon_\kk^- \left( 1 + 2N(\epsilon_\kk^-) \right) 
- 2 E_\kk
\right),
\end{equation}
where $N(\omega)=1/(e^{\omega/T}-1)$, and which can be expressed using two quantities in the leading order of small parameters
\begin{eqnarray}
\label{f1}
	f_1 &=& 
	\frac1N \sum_{\bf k} 
	\left(\frac{J_{\bf 0}}{2\sqrt{J_\0^2-J_\kk^2}} \left( 1 + N(\epsilon_\kk^+) + N(\epsilon_\kk^-) \right)  - \frac12\right),\\
\label{f2}
	f_2 &=& \frac1N \sum_{\bf k} \frac{J_{\bf k}^2}{2J_\0\sqrt{J_\0^2-J_\kk^2}}\left( 1 + N(\epsilon_\kk^+) + N(\epsilon_\kk^-)\right).
\end{eqnarray}
Numerical integration in Eqs.~\eqref{f1} and \eqref{f2} gives $f_1	\approx0.1966$ and $f_2	\approx 0.2756$ at $T=0$. In particular, the renormalized spin value $\langle S\rangle$ is given by $\langle S\rangle = S-f_1$. Counterparts of Eqs.~\eqref{eclcol} and \eqref{eclsf} in the first order in $1/S$ have the form (up to terms which are common to both energies)
\begin{eqnarray}
\label{eclcolq}
\frac1N E^{col} &=& \frac{S^2}{2} \left(J'_\kol - Q_\kol^{bb}\right)\left(1-\frac2Sf_1\right), \\
\label{eclsfq}
\frac1N E^{sf} &\approx& \frac{S^2}{2} \left(J'_\kog - Q_\kog^{cc}\right)\left(1-\frac2Sf_1\right) - \frac{H^2}{4J_\0}\left(1-\frac1Sf_2\right),
\end{eqnarray}
Comparing Eqs.~\eqref{eclcolq} and \eqref{eclsfq}, one obtains for the spin-flop field
\begin{equation}
\label{hsfq}
H_{sf} = \left(S - f_1 + \frac12f_2\right) \sqrt{ 2J_\0 \left( 8(J_{ac}'-J_{ab}') + Q^{bb}_\kol - Q^{cc}_\kog \right) }
\end{equation}
which differs a factor 0.87 from its classical counterpart \eqref{hsf} at $S=1/2$ and $T=0$.

It can be shown that inequality \eqref{ineq} and Eq.~\eqref{psic} for $\psi_c$ are not renormalized by fluctuations in the first order in $1/S$ because all terms (except for those proportional to $J_\0$ and $H^2$) in Eqs.~\eqref{eclsf} and \eqref{eclsfin} acquire the same factor $1-2f_1/S$ (cf.~Eqs.~\eqref{eclcolq} and \eqref{eclsfq}). Terms not proportional to $J_\0$ and $H^2$ are renormalized in the same way in Eqs.~\eqref{ea} and \eqref{eb}. Besides, fluctuations lead also to extra terms $\sin^2\xi S\delta_\0f_2/2$ and $S\delta_\0f_2/2$ in Eqs.~\eqref{ea} and \eqref{eb}, respectively, which lead to the following renormalization of the critical angle $\xi_c$ in the first order in $1/S$ (cf.~Eq.~\eqref{xic}):
\begin{equation}
\label{xicr}
\cos\xi_c = \sqrt{
\frac{8 (J_{ac}-J_{ab})-Q_\kog^{cc} + Q_\kol^{bb}}{(1-f_2/S) \delta_\0 - Q_\kol^{xx} + Q_\kol^{bb} }
}.
\end{equation}

\begin{figure}
\noindent
\includegraphics[scale=0.4]{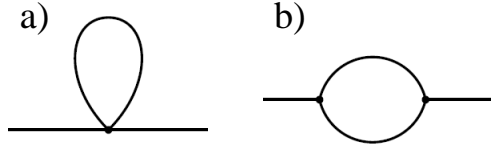}
\hfil
\caption{Diagrams of the first order in $1/S$ for self-energy parts discussed in the present paper. Diagrams (a) and (b) come from four- and three-magnon terms in the Hamiltonian, respectively.}
\label{dia}
\end{figure}

Renormalization of magnon spectrum stems in the first order in $1/S$ from diagrams shown in Fig.~\ref{dia}. One obtains after straightforward calculations (see, e.g., Ref.~\cite{batalov}) that $\omega_{1,2}^{col}(H)$ are given by Eqs.~\eqref{o1} and \eqref{o2}, where now (cf.\ Eqs.~\eqref{o10} and \eqref{o20})
\begin{eqnarray}
\label{o1colq}
\omega_1^{col}(0) &=& \left(S - f_1 + \frac12f_2\right)\sqrt{2J_\0 \left( Q^{bb}_\kol - Q^{cc}_\kol \right)} ,\\
\label{o2colq}
\omega_2^{col}(0) &=& (S-f_1)\sqrt{2J_\0 \delta_\0}.
\end{eqnarray}
Eq.~\eqref{o1sf} give $\omega_1^{sf}(H)$, where now (cf.\ Eq.~\eqref{o1sf0})
\begin{equation}
\label{o1sfq}
\tilde\omega_1 = \left(S - f_1 + \frac12f_2\right)\sqrt{ 2J_\0 \left( Q^{cc}_\kog - Q^{bb}_\kog \right) },
\end{equation}
and one has for $\omega_2^{sf}$ instead of Eq.~\eqref{o2sf}
\begin{equation}
\label{o2sfq}
\omega_2^{sf}(H) = \epsilon_{\kk=\0}^+ =  (S-f_1)\sqrt{2J_\0 \delta_\0} = \omega_2^{col}(0).
\end{equation}

The gap in the lower magnon branch is given by Eqs.~\eqref{gapcol} and \eqref{gapsf}, where $H_{sf}$ is the renormalized spin-flop field given by Eq.~\eqref{hsfq}. 

\section{Comparison with experiment}
\label{compexp}

Magnon spectrum analysis carried out in $\rm Cu(pz)_2(ClO_4)_2$ in neutron scattering experiment \cite{pz3} gives $J\approx18.1(4)$~K. The value of magnetic moment was found at $T=2.3$~K and $H=0$ to be $0.47(5)\mu_B$ that corresponds to the renormalized spin value of $\langle S\rangle = 0.23(2)$. In accordance with the experiment, one obtains at $T=2.3$~K using Eq.~\eqref{f1} $\langle S\rangle = S-f_1\approx0.24$ (notice that thermal corrections move theoretical result closer to the experimentally obtained value: one finds using Eq.~\eqref{f1} $\langle S\rangle = S-f_1\approx0.3$ at $T=0$).

The spin-flop field value was obtained experimentally to be 0.44(3)~T at $T=1.3$~K. \cite{smirnov,priv} One finds using this value and Eq.~\eqref{hsfq} that $J_{ac}'-J_{ab}'$ approaches the upper limit determined by Eq.~\eqref{ineq}:
\begin{equation}
\label{jpp}
J_{ac}'-J_{ab}' \approx 0.0016(3)~{\rm K}.
\end{equation}
This value (as well as the upper limit for $J_{ac}'-J_{ab}'$ in Eq.~\eqref{ineq}) is an order of magnitude smaller than the existing estimation of $\sim10^{-2}$~K for the inter-plane exchange coupling value proposed before (see Ref.~\cite{fam} and references therein). However we point out that dipolar forces were not taken into account in previous estimations of inter-plane coupling. On the other hand, the value of the inter-plane dipolar coupling is of the order of $10^{-2}$~K as it is seen, e.g., from Eqs.~\eqref{qq}.

Result is presented in Fig.~\ref{expfig} of the fit of AFR data obtained in Ref.~\cite{smirnov} at $T=1.3$~K using formulas for $\omega_{1,2}(H)$ from Sec.~\ref{qcor}. In particular, we find for the easy-plane anisotropy using expressions for $\omega_2^{col}(H)$ (i.e., Eqs.~\eqref{o2}, \eqref{f1}, \eqref{f2}, and \eqref{o2colq})
\begin{equation}
\label{dval}
	\delta/J \approx 3.4\times 10^{-3}
\end{equation}
that is in agreement with values of $2\times 10^{-3}$, $3\times 10^{-3}$, $5\times 10^{-3}$, and $5.5\times 10^{-3}$ reported for this quantity in Refs.~\cite{pz3,smirnov,fam1,fam}, respectively. To plot $\omega_1(H)$ at $H>H_{sf}$ in Fig.~\ref{expfig}, we use Eqs.~\eqref{o1sf} and \eqref{o1sfq}, which do not contain fitting parameters. Eqs.~\eqref{o1} and \eqref{o1colq} are derived above for $\omega_1(H)$ at $H<H_{sf}$ in the leading order in small parameters (in particular, in the leading order in $\omega_0/\delta_\0$). We have found, however, that contribution to Eqs.~\eqref{o1} and \eqref{o1colq} of higher-order terms is small but quite noticeable at $H\approx H_{sf}$ in $\rm Cu(pz)_2(ClO_4)_2$. This contribution moves theoretical results closer to experimental data. Then, to plot $\omega_1(H)$ at $H<H_{sf}$ in Fig.~\ref{expfig}, we use Eqs.~\eqref{dval} and \eqref{o1}, where
\begin{equation}
\omega_1^{col}(0)^2 = 
2S(S - 2f_1 + f_2) J_\0 \left( Q^{bb}_\kol - Q^{cc}_\kol \right)
+
\frac{2H^2}{S^3J_\0\delta_\0} 
\left( 
H^2(S+2f_1) - 2S^2(S+f_2)J_\0 \left( Q^{bb}_\kol - Q^{cc}_\kol \right) 
\right).
\end{equation}
The first term in this expression corresponds to Eq.~\eqref{o1colq} whereas the last one is of the next order in $\omega_0/\delta_\0$ at $H\approx H_{sf}\sim S\sqrt{J_\0\omega_0}$ compared to the first term. 
\footnote{
Notice that terms proportional to $\omega_0$ are accompanied by numerically small factors of the order of 0.1 (see Eqs.~\eqref{qq}). This makes valid the expansion in parameter $\omega_0/\delta_\0$ in $\rm Cu(pz)_2(ClO_4)_2$.
}

One obtains from Eqs.~\eqref{psic} and \eqref{jpp} $\psi_c=20(6)^\circ$ that is close (taking into account the error) to $10^\circ$ found in Ref.~\cite{smirnov}. Eqs.~\eqref{xicr} and \eqref{dval} give $\xi_c\approx60^\circ$ that is twice as large as the experimental result of $30^\circ$. \cite{smirnov} 

We point out strong quantum fluctuations as the most probable source of discrepancies between the theory and experiment in $\omega_1(H)$ at $H<H_{sf}$ (see Fig.~\ref{expfig}) and in the critical angle $\xi_c$. Really, corrections of the first order in $1/S$ taken into account above renormalize strongly classical values of observables ($f_1\approx0.22$ and $f_2\approx0.3$ in $\rm Cu(pz)_2(ClO_4)_2$ at $T=1.3$~K). In particular, the factor $1-f_2/S$ is equal approximately to 0.4 in Eq.~\eqref{xic}. On the other hand, a small increasing of $f_2$ to approximately 0.4 results in $\xi_c\approx35^\circ$ that is much closer to the experimentally found value of $30^\circ$. Then, it seems to us likely that further-order $1/S$-corrections can improve the quantitative agreement with experiment (even if these corrections are smaller than the first-order ones).
\footnote{
It is well known that higher-orders $1/S$-corrections are small and the first $1/S$ corrections give the major contribution to renormalization of such observable quantities as staggered magnetization, spin-wave velocity, transverse susceptibility, and some others even in spin-$\frac12$ 2D AFs. \cite{monous} However, examples of observable quantities having badly converged $1/S$ series are also well-known in 2D AFs with $S\sim1$: short-wavelength magnon spectrum \cite{af2d,syromyat2,zhitchern} and chiral dynamical susceptibility \cite{syromyat}.} 
However corresponding detailed consideration is out of the scope of the present paper.

\section{Summary and conclusion}
\label{conc}

To summarize, we discuss ground state properties and magnon spectrum in the first order in $1/S$ of model \eqref{ham} describing layered Heisenberg AF with small easy-plane anisotropy. We obtain that this model shows all essential features obtained so far experimentally in $\rm Cu(pz)_2(ClO_4)_2$. We demonstrate that dipolar forces lead to the in-plane easy-axis anisotropy observed experimentally due to small departure of the crystal structure from tetragonal one. The spin-flop field value is given by Eq.~\eqref{hsfq}. A peculiar characteristic feature of dipolar interaction is that the magnetic ordering changes upon the spin-flop transition in the direction transverse to square planes if inter-plane exchange coupling is sufficiently small: vector of magnetic structure $\ko$ changes according to Eq.~\eqref{ko} if inequality \eqref{ineq} holds. This changing of the magnetic ordering takes place at $\xi<\xi_c$ and $\psi<\psi_c$, where critical angles $\xi_c$ and $\psi_c$ are given by Eqs.~\eqref{xicr} and \eqref{psic} in the first order in $1/S$, respectively (see inset in Fig.~\ref{expfig}). We demonstrate that the magnetic ordering changing leads also to quite an unusual characteristic behavior of AFR frequencies presented in Fig.~\ref{expfig} and given by Eqs.~\eqref{o1}, \eqref{o2}, \eqref{o1colq}, \eqref{o2colq} and Eqs.~\eqref{o1sf}, \eqref{o1sfq}, \eqref{o2sfq}, at $H<H_{sf}$ and $H>H_{sf}$, respectively. The peculiar behavior of lower ($\omega_1$) AFR frequency is related to the fact that it is not equal to the magnon gap. The latter circumstance in turn is a consequence of the switching \eqref{ko} of $\ko$ at the spin-flop transition. As usual, the magnon gap vanishes at $H=H_{sf}$ and it is given by Eqs.~\eqref{gapcol} and \eqref{gapsf} at $H<H_{sf}$ and $H>H_{sf}$, respectively (see also Fig.~\ref{expfig}).

Fitting parameters of model \eqref{ham}, we reach a good quantitative agreement with almost all available low-temperature experimental data obtained so far in $\rm Cu(pz)_2(ClO_4)_2$. We expect, however, that one has to go beyond the first order in $1/S$ to reach better quantitative agreement between the theory and experiment. This is particularly true for $\xi_c$ due to high sensitivity of this quantity to values of quantum and thermal corrections.

Our results can be relevant to a family of recently synthesized 2D spin-$\frac12$ Heisenberg AFs to which $\rm Cu(pz)_2(ClO_4)_2$ is a prototype. \cite{fam} 

Finally, we point out that the effect of dipolar forces (leading to the in-plane easy-axis anisotropy, to the switching \eqref{ko} of $\ko$, and to peculiar behavior of AFR frequencies) can be described phenomenologically by the following model with only short-range spin interaction:
\begin{equation}
\label{hameff}
\mathcal{H}_{eff} = \sum_{\langle l, m\rangle}
\left( J{\bf S}_l {\bf S}_{m}
+
A S_l^b S_{m}^b
-
\delta S_l^x S_{m}^x \right)
+
\sum_{\langle l, m\rangle}
\left( J'_{lm}{\bf S}_l {\bf S}_{m} - B_{lm} S_l^c S_{m}^c \right)
-
{\bf H}\sum_l {\bf S}_l,
\end{equation}
where $\langle l, m\rangle$ in the first and in the second terms denote couples of nearest-neighbor spins from the same and from the neighboring $bc$ planes, respectively, $A,\delta>0$, and $B_{lm}=B>0$ and $B_{lm}=0$ for spins lying in the same $ac$ and $ab$ planes, correspondingly. Compared to model \eqref{ham}, Eq.~\eqref{hameff} does not contain dipolar interaction and two anisotropic coupling $A$ and $B_{lm}$ are added. One easily infers that if $J\gg \delta\gg A\gg B\gg J'_{ac}-J'_{ab}>0$, $b$ is an easy axis and $\ko=\kol$ at $H=0$. On the other hand, $\ko$ switches from $\kol$ to $\kog$ at the spin-flop transition because the anisotropic interaction $B_{lm}$ dominates over the inter-plane exchange coupling $J'$. As a result, it can be shown that model \eqref{hameff} has the same peculiar characteristic features as model \eqref{ham} which are discussed above. Thus, the effect of dipolar interaction in the microscopic model \eqref{ham} is simulated by the hierarchy of anisotropic interactions in the phenomenological model \eqref{hameff}.

\begin{acknowledgments}

I thank A.\ I.\ Smirnov and K.\ Yu.\ Povarov for useful discussions of experimental results and exchange of data. This work is supported by Russian Science Foundation (grant No.\ 14-22-00281).

\end{acknowledgments}

\bibliography{BibAnisChange}

\end{document}